# Bloch Oscillations of Einstein-Podolsky-Rosen States

Maxime Lebugle, Markus Gräfe, René Heilmann, Armando Perez-Leija, Stefan Nolte, and Alexander Szameit[*]

Institute of Applied Physics, Abbe Center of Photonics, Friedrich-Schiller-Universität Jena, Max-Wien-Platz 1, 07743 Jena, Germany.

* Corresponding author: alexander.szameit@uni-jena.de

**Bloch Oscillations (BOs) of quantum particles manifest themselves as periodic spreading and re-localization of the associated wave functions when traversing lattice potentials subject to external gradient forces[1]. Albeit BOs are deeply rooted into the very foundations of quantum mechanics, all experimental observations of this phenomenon so far have only contemplated dynamics of one or two particles initially prepared in separable local states, which is well described by classical wave physics. Evidently, a more general description of genuinely quantum BOs will be achieved upon excitation of a Bloch-oscillator lattice system by nonlocal states[2], that is, containing correlations in contradiction with local realism[3]. Here we report the first experimental observation of BOs of two-particle Einstein-Podolsky-Rosen states (EPR), whose associated $N$-particle wave functions are nonlocal by nature. The time evolution of two-photon EPR states in Bloch-oscillators, whether symmetric, antisymmetric or partially symmetric, reveals unexpected transitions from particle antibunching to bunching. Consequently, the initial state can be tailored to produce spatial correlations akin to bosons, fermions or anyons[4]. These results pave the way for a wider class of photonic quantum simulators[5-7].**

In his pioneering work[1], Felix Bloch predicted the existence of periodic spatial oscillations of electrons traversing crystalline structures driven by linear external forces. At that time, understanding the concept of BOs contributed to the establishment of the electronic band structure theory in solids, which later on impacted wide areas of physics. In fact, BOs are a universal wave phenomenon and have since been formally investigated using diverse physical platforms[8-10]. Thus, while the observation of BOs is still elusive in its original conceived setting, a formal analogy between the propagation of classical beams and the evolution of point-like quantum particles can be appreciated. It has led to experimental realizations of photonic BOs in arrays of evanescently coupled waveguides[11-13].

The aim of the present work is to investigate BOs of two-particle EPR states. As it will be elucidated, our experiments help unravel the time-dynamics of quantum entanglement through Bloch-oscillator lattices and allow observation of nonclassical correlations that persist along the propagation[2,14]. We demonstrate that, as a result of spatial entanglement and quantum coherence, the arising multi-particle interference strongly influences the particle statistics exhibiting periodic Hanbury Brown-Twiss bunching/antibunching processes[14-16]. Furthermore, we also show that the symmetry imposed on the input state enriches the observations yielding to some intriguing quantum statistics allusive to the behavior of bosons, fermions, and anyons. In that sense, we realize a versatile quantum simulator[5-7] able to simulate the dynamics of three particle families within a single scheme.

Our approach is based on integrated photonic circuits, a propitious platform to study fundamental quantum phenomena such as random walks[17-19] as well as possible building blocks for quantum photonic chips[20,21]. The monolithic nature of such integrated devices gives access to full interferometric control over the field dynamics. As such, they are ideal candidates for the realization of quantum simulators[22-25] capable to explore physical processes beyond the classical limits. Moreover, using photon as the carrier of information makes it feasible to define EPR states on the spatial degree of freedom; since therein no particle-particle interaction can hinder the observation of subtle interference effects.

To study the quantum dynamics of a tight-binding Bloch-oscillator lattice[13,14], we consider the Heisenberg equations of motion for field operators $a_k^\dagger$ in the $k$-th site

$$i\frac{da_k^\dagger}{dz} = kBa_k^\dagger + C\left(a_{k+1}^\dagger + a_{k-1}^\dagger\right), \qquad (1)$$

where $z$ is the spatial coordinate mapping the time variable, $C$ is the inter-mode coupling coefficient, and $B$ denotes the difference in the propagation constants of adjacent modes (Fig. 1a). Integration of equation (1) yields the evolution of the creation operators $a_k^\dagger(z) = \sum_m U_{k,m}(z) a_m^\dagger(0)$ with $U_{k,m}(z) = \left[\exp(-iz\mathrm{H})\right]_{k,m}$ representing the evolution matrix, and H the coupling coefficient matrix. The Bloch period is defined with $\lambda_B = 2\pi/B$ as the distance at which the first revival is expected. When using EPR biphoton states (Fig. 1b), amplitude interference effects occur in the Bloch-oscillator, which arise from the complex superposition of paths the pair may take. In terms of photon density, a substantial broadening of the wave function is observed up to half of the Bloch cycle, further followed by a re-localization over $\lambda_B$ (Fig. 1c, left inset). On the other hand, two-particle correlation analysis reveals that the presence of the ramping potential forces the photons to separate and gather cyclically (Fig. 1c, right inset). It causes bunching to antibunching transitions, or vice versa, in the vicinity of some particular distances which we will refer to as correlation turning points throughout our manuscript. We have implemented Bloch arrays of $N=16$ single-mode optical waveguides (Fig. 2b) using the femtosecond-laser writing approach[26] in glass. All the waveguides have identical propagation constants $\beta_0$ and are separated by $30\,\mu\mathrm{m}$ such that the coupling coefficients are equal. Here, the effective linear ramping potential required to induce BOs is synthetically obtained by bending the waveguides[27].

Ideal two-photon EPR states on a discrete photonic lattice are defined as

$$|\Psi_{EPR}\rangle = \frac{1}{\sqrt{2}}\left[|2_m, 0_n\rangle + \mathrm{e}^{i\phi}|0_m, 2_n\rangle\right], \qquad (2)$$

where the indices refer to the lattice sites and $\phi$ is a relative phase. The quantum correlations in this nonlocal $|\Psi_{EPR}\rangle$ state are such that the two photons always couple into the same excited lattice site, either $m$ or $n$, with an equal probability $1/2$. For our experiments, we consider three different two-photon EPR input states in adjacent modes, namely with symmetric ($\phi=0$), antisymmetric ($\phi=\pi$), and partially symmetric exchange amplitudes with $\phi \in (0,\pi)$. Importantly, all three states are prepared on-chip which guarantees ultra-high experimental stability. We achieve this by introducing detuned directional couplers (Fig. 2c, see Methods).

First, we investigate the time evolution of symmetric two-photon EPR states $|\Psi_S\rangle$ in the system. To do so, $815-$nm photon pairs with horizontal polarization were generated by type-I parametric down-conversion and coupled into either channel zero or one of the Bloch-oscillator array with the same probability, $|\Psi_S\rangle = \left[\left(a_0^\dagger\right)^2 + \left(a_1^\dagger\right)^2\right]|0\rangle/2$. In order to monitor the time-evolution of quantum correlations, we fabricated several curved arrays of the same physical length ($6\,\text{cm}$) but different Bloch periods $\lambda_B$. By tuning the radius of curvature (Fig. 2d), we have specifically implemented five arrays for each input state effectively achieving fractions of the Bloch periods $0.1\lambda_B$, $0.2\lambda_B$, $0.3\lambda_B$, $0.4\lambda_B$ and $0.5\lambda_B$ (see Methods).

Once the $|\Psi_S\rangle$ state is coupled into the system, the spatial correlation function is obtained through measurements of the probability of simultaneously detecting one photon exiting from waveguide $k$ and its twin at site $l$ (Fig. 2a). Mathematically, this is described[14] by $\Gamma_{k,l}(z) = \langle a_k^\dagger a_l^\dagger a_l a_k \rangle$. Figure 3a-d shows the correlation matrices obtained after the propagation distances of $0.1\lambda_B \ldots 0.4\lambda_B$, respectively. At $z = 0.1\lambda_B$, the correlation map clearly exhibits that the photons initially correlated in their positions have undergone a sudden split-up, becoming anticorrelated (Fig. 3a). Subsequently, after propagating over $z = 0.2\lambda_B$, $\Gamma_{k,l}(z)$ reveals a correlation turning point with co-occurrence of correlation and anticorrelation among the constitutive photons (Fig. 3b). This regime is a clear manifestation of a mixed quantum statistics resembling the one expected from anyonic particles - quantum entities that are neither fermions nor bosons ruled by fractional exclusion statistics[4]. After that point the state becomes once more correlated (Fig. 3c). These bunching effects turn out to be evident as the particles approach half of the Bloch cycle (Fig. 3d). Our results are fully supported by best-fit simulations obtained with input phase shift of $\phi_S^{\text{exp}} = 0.1\pi \pm 0.1\pi$ (shown beside in Fig. 3a-d). This small discrepancy when compared to a perfect state can be explained by the presence of losses in the directional coupler section[28].

To gain more insights into the correlation patterns, we extracted from the matrices the interparticle distance probability given by $g(\Delta) = \frac{1}{\delta}\sum_q \Gamma_{q,q+\Delta}$, where $\delta$ is the number of combinations $\Gamma_{q,q+\Delta}$ satisfying the inequality $q+\Delta \leq N$. Figure 4a shows how $g(\Delta)$ evolves for the symmetric input. We observe a gradual transformation in $g(\Delta)$ upon propagation, from a double-peak structure to a single-peak one, which confirms that the photon pair passes a correlation turning point. Furthermore, to evidence probability interference as the underlying mechanism of the transition, we performed additional measurements using the device at $0.4\lambda_B$ with two distinguishable photons which thereby manifest no quantum interference (see Extended Data Figure 1 and Supplementary Information).

To analyze the impact of the input symmetry property over the photon correlations, as a second case we consider the propagation dynamics of an antisymmetric state, $|\Psi_{AS}\rangle = \left[\left(a_0^\dagger\right)^2 - \left(a_1^\dagger\right)^2\right]|0\rangle/2$. Remarkably, it presents similar characteristics upon propagation (Fig. 3e-h). However, the initial $\pi$-phase shift among the modes offsets the correlation cycle to start from a clear bunched state at $0.1\lambda_B$ (Fig. 3e). A singular transition again occurs between the distances $0.2\lambda_B$ and $0.3\lambda_B$, resulting in a strongly anticorrelated state at $0.4\lambda_B$ (Fig. 3h). Extracting the interparticle distance probability also provides a clear picture of the progressive transformation from correlation to anticorrelation (Fig. 4b). Best-fit simulations are obtained with $\phi_{AS}^{\exp} = 1.0\pi \pm 0.1\pi$ (Fig. 3e-h), and demonstrate the excellent control of the antisymmetric state preparation. For all cases, we measure the correlation similarity between theoretical predictions and our experimental observations, $S = \left(\sum_{k,l}\sqrt{\Gamma_{k,l}^{exp}\Gamma_{k,l}^{th}}\right)^2 / \left(\sum_{k,l}\Gamma_{k,l}^{exp}\sum_{k,l}\Gamma_{k,l}^{th}\right)$. This parameter lies between $S = 0.914 \pm 0.001$ and $S = 0.952 \pm 0.002$, which indicates the high performances of our devices (Extended Data Table 1).

In order to get a whole picture of the influence of the initial state symmetry on the correlation dynamics, as a third case we focus on the evolution of partially symmetric two-photon EPR states. Such photonic states are known to produce fractional quantum statistics ($a_m^\dagger a_n^\dagger = e^{i\phi} a_n^\dagger a_m^\dagger$, $0 < \phi < \pi$) and therefore are able to emulate the behavior of anyonic particles, which present features of both Bose-Einstein and Fermi-Dirac statistics. For best observation of anyonic-like statistics, we prepared the arbitrary state $|\Psi_{arb.}\rangle = \left[ (a_0^\dagger)^2 + e^{i\phi_{arb.}^{exp.}} (a_1^\dagger)^2 \right] |0\rangle / 2$, with $\phi_{arb.}^{exp.} = 0.8\pi \pm 0.1\pi$. Extended Data Figure 2 plots the correlation map obtained at fixed propagation distance of $0.3\lambda_B$ along with those obtained for symmetric and antisymmetric state preparation. As anticipated, tuning the input phase shift offsets the correlation cycle. It results in a concurrence of photon bunching and antibunching for the $|\Psi_{arb.}\rangle$ state input case, and demonstrates the control achieved over the quantum statistics.

In our Bloch system, the linear potential continuously induces position-dependent phase shifts on each mode, which has a dramatic impact on the correlation dynamics. More specifically, quantum states propagating in the lattice experience enhancement or reduction of correlation probabilities depending on the position they are passing through. Hence, it accounts for this peculiar form of spatially extended multi-path Hong-Ou-Mandel effect, generating oscillations between two diagonal and two off-diagonal peaks in the correlation matrix. Besides, it reflects the intrinsic capability of the lattice to efficiently tailor the associated quantum particle statistics; e.g., for the symmetric input from fermionic-like at short propagation distances over anyonic- to bosonic-like at half the Bloch cycle. In this vein, the quantum simulator realized here, when seeded with two-photon EPR states, hence permits the simulation of three different types of particles.

We finally estimate the nonclassicality of the states. To do so, the Bell-like inequality $V_{k,l} = \frac{2}{3}\sqrt{\Gamma_{k,k}\Gamma_{l,l}} - \Gamma_{k,l} < 0$ is employed[14]. Its violation provides a valid criterion to discern quantum correlations that manifest themselves as photon bunching. Exemplarily, violations for the $|\Psi_S\rangle$ state at $z = 0.4\lambda_B$ and for the $|\Psi_{AS}\rangle$ state at $z = 0.1\lambda_B$ are shown in Fig. 4c, d. The maximum positive value reaches $65\sigma_{k,l}$, where the standard deviation $\sigma_{k,l}$ is determined assuming Poissonian statistics for each count. With distinguishable photon pairs, the inequality is never violated experimentally, proving that the correlations are purely quantum.

We have experimentally investigated Bloch oscillations of two-photon EPR states. This system exhibits singular transitions between particle bunching and antibunching, thereby endowing photonic simulators with the ability to gradually switch a given particle quantum statistics from bosonic- over anyonic- to fermionic-like. Our work sheds light on unique nonlocal features of two-photon EPR states and demonstrates the high level of nonclassicality currently observable in photonic circuits. In this vein, our implementation of path-entangled states with fractional exchange statistics could possibly find applications in topological quantum computing, and more generally in mastering quantum solid-state processes for future computation schemes.

## Methods

**Devices fabrication and design.** The waveguides were written inside high-purity fused silica (Corning 7980, ArF grade) using a RegA 9000 seeded by a Mira Ti:Al$_2$O$_3$ femtosecond laser. Pulses centered at 800 nm with duration of 150 fs were used at a repetition rate of 100 kHz and energy of 450 nJ. The pulses were focused 250 $\mu$m under the sample surface using a NA = 0.6 objective while the sample was translated at constant speed of 60 mm·min$^{-1}$ (except for the detuned sections) by high-precision positioning stages (ALS130, Aerotech Inc.). The mode field diameters of the guided mode were 18 $\mu$m × 20 $\mu$m at 815 nm. At the wavelength of interest, propagation losses and birefringence were estimated at 0.3 dB·cm$^{-1}$ and in the order of $10^{-7}$, respectively. The waveguides are equally spaced by 127 $\mu$m at the two facets in order to match standard V-groove fiber arrays for in- and out-coupling of single photons. The waveguides then smoothly converge through fanning arrangements to their eventual configuration in the functional sections.

**Experimental set-up.** Pairs of $815-\text{nm}$ photons were produced by type-I optical spontaneous parametric down-conversion from a BiB$_3$O$_6$ nonlinear crystal pumped with 70 mW from a continuous-wave diode laser, and further filtered by $3-\text{nm}$ interference filters to increase the photon indistinguishability. We successively measured the signals exiting from the even and odd modes, which are collected via a butt-coupled fiber array. They were further coupled to their respective avalanche photodiodes (Fig. 2a) connected to a photon correlator card (Becker & Hickl: DPC230). Data analysis of the channel counts was realized with a time window set at $1\,\text{ns}$. Accidental coincidences, that is, simultaneous detection of two photons not coming from the same pair is estimated to occur with a negligible rate of less than $2\times 10^{-6}$ per second. The measurements are acquired over 15 minutes of integration time and are corrected for coupling fluctuations using simultaneously detected single-photon probability distribution. Each channel is corrected for relative detection efficiencies.

**Arbitrary EPR state preparation.** Two-photon EPR states can be created by simultaneously exciting the two modes of an integrated $50:50$ Directional Coupler (DC) with indistinguishable photons in a separable product state[29,30]. More importantly, the underlying flexibility of this arrangement can be used to conceive a diverse family of two-photon states by adjusting the relative phase $\phi$ in the range $\phi \in (0,\pi)$ between the associated propagating modes (Fig. 2c). We accomplish this by inducing an extra detuning $\Delta\beta$ of the propagation constant $\beta_0$ in one of the arms of the DC, introducing so-called detuned DCs (see Supplementary Information). During the fabrication process, the control over $\Delta\beta$ is readily achieved through adjustment of the writing speed, hence giving access to arbitrary phase shifts.

**On-chip Bloch oscillations.** Observing the time-dynamics in the Bloch-oscillator is made possible by fabricating several devices with different radius of curvature. Indeed, the ramping parameter $B$ directly depends on the radius of curvature $R_C$ of the waveguides according to the expression $B=\Omega/R_C$ where $\Omega$ is equal to $2\pi n_{eff}d/\lambda$, $n_{eff}$ is the effective refractive index of the waveguides, $d$ the inter-waveguide separation and $\lambda$ the wavelength[27]. As a result, the ramping parameter can be adjusted by tuning $R_C$, thus affecting the Bloch period. This strategy allows for changes in the effective propagation length while keeping the broadening of the wave function closely comparable between all cases.

**Acknowledgments**

This work was supported by the Marie Curie Actions within the Seventh Framework Programme for Research of the European Commission, under the Initial Training Network PICQUE, Grant No. 608062. The authors further gratefully acknowledge financial support from the German Ministry of Education and Research (Center for Innovation Competence program, Grant No. 03Z1HN31), the Thuringian Ministry for Education, Science and Culture (Research group Spacetime, Grant No. 11027-514), and the Deutsche Forschungsgemeinschaft (Grant No. NO462/6-1).

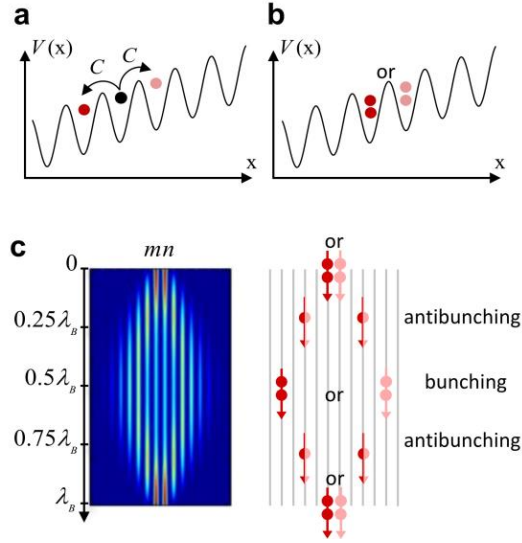

**Fig. 1.** Quantum Bloch-oscillator. **a,** Bloch oscillations of a single particle when subject to a periodic potential $V(x)$ with an external constant force acting on it, resulting in state entanglement (displayed as different colors). The coupling coefficient is denoted as $C$. **b,** Bloch oscillations of the state $|\Psi_S\rangle = \left[(a_m^\dagger)^2 + (a_n^\dagger)^2\right]|0\rangle/2$ input in adjacent modes, implying probability interference that may either be constructive or destructive. **c,** Response of a Bloch-oscillator initially excited with the $|\Psi_S\rangle$ state. The particle density shows a perfect revival over one Bloch period $\lambda_B$ (left inset). The corresponding correlation dynamics are schematically presented in the right inset. It conceptualizes the bunching to antibunching transitions governing the state all over the cycle.

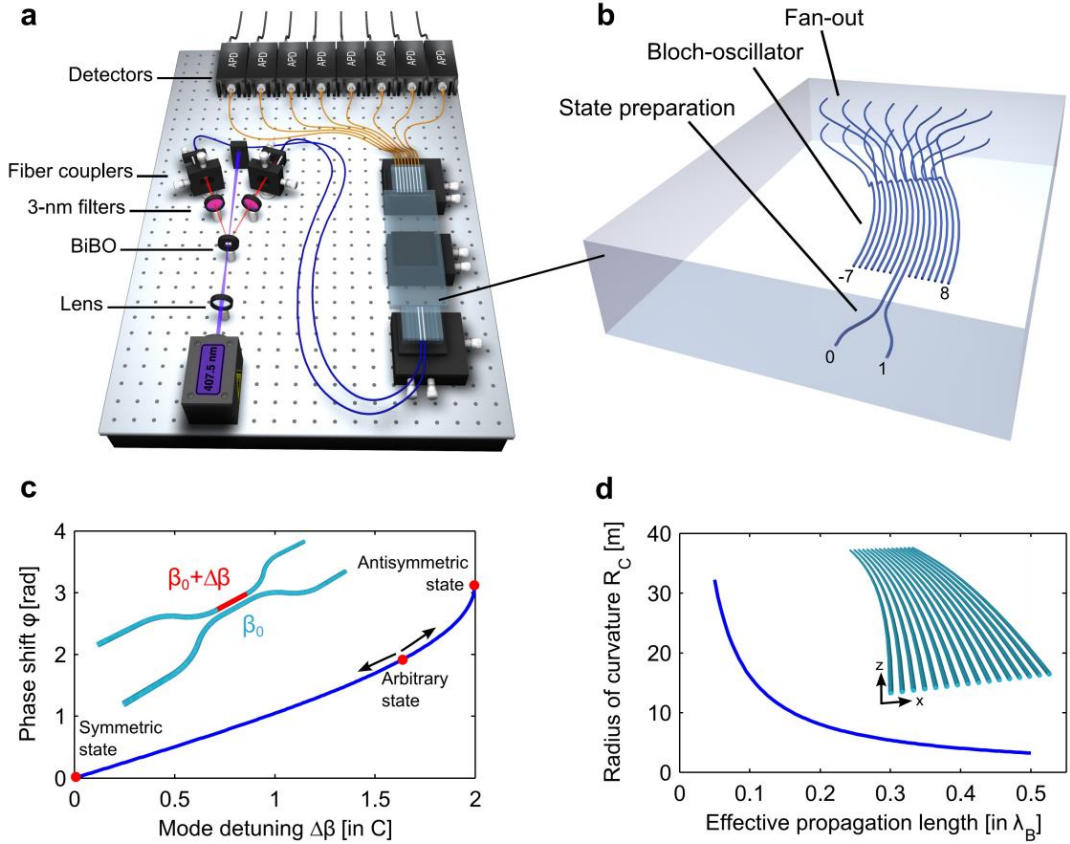

**Fig. 2. a,** Spatial correlations measurement set-up. Photon pairs are produced by type-I spontaneous parametric down-conversion and single-photon detection is achieved using avalanche photodiodes. **b,** Femtosecond-laser-written photonic circuit made of two functional sections: A two-photon EPR state with arbitrary symmetry is first prepared in a detuned directional coupler, and subsequently evolves in a Bloch-oscillator emulated by a curved array. **c,** Preparation of two-photon EPR state with arbitrary phase shift $\phi$, achieved by detuning the propagation constant by $\Delta\beta$ in one of the two arms of a $50:50$ directional coupler. **d,** Radius of curvature $R_C$ of waveguides arranged in a curved array necessary for reaching a specific point along the Bloch's cycle.

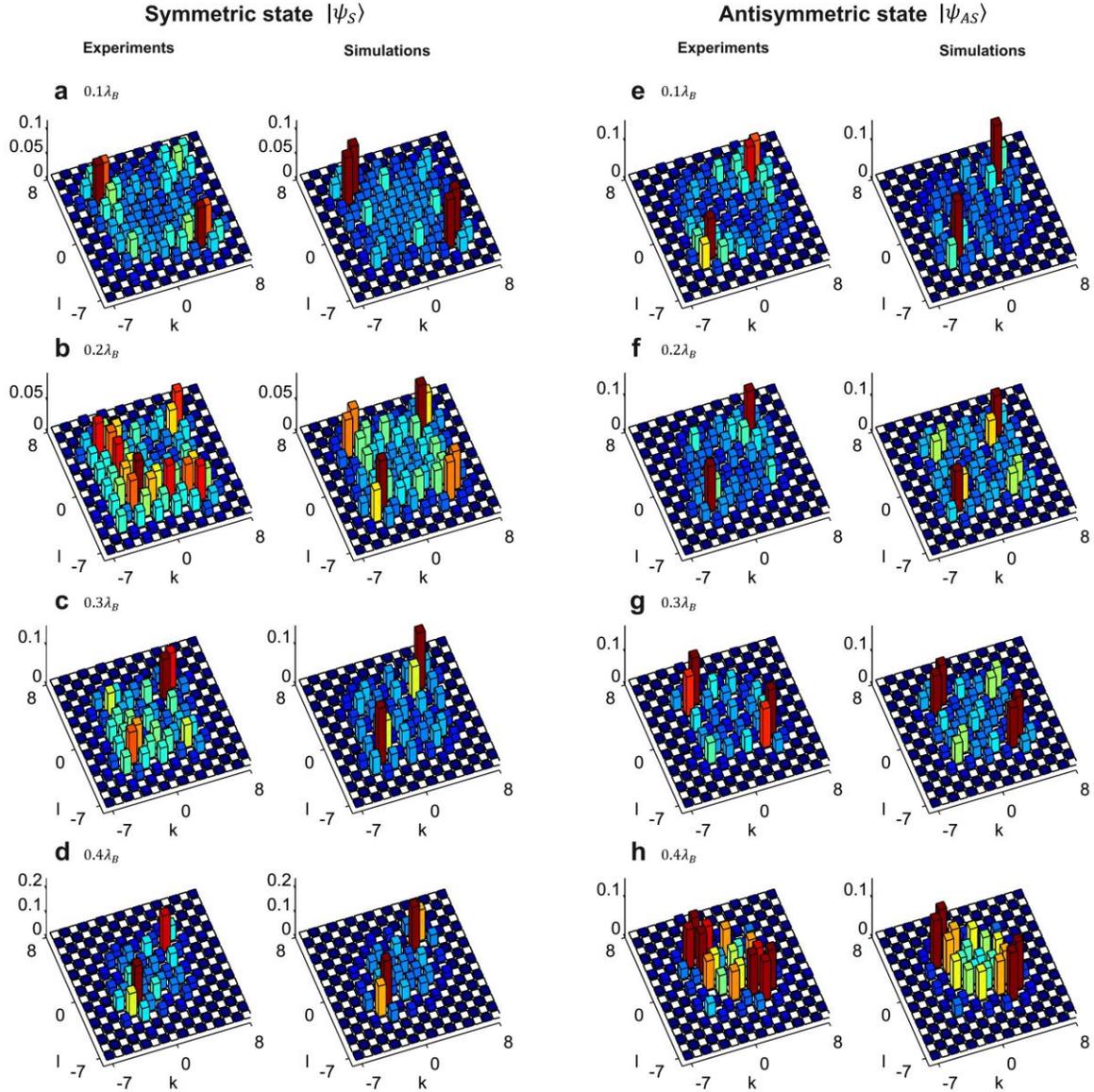

**Fig. 3.** Experimental correlation matrices $\Gamma_{k,l}(z)$ for four discrete propagation lengths $0.1\lambda_B \ldots 0.4\lambda_B$ obtained with **a-d** a symmetric input state $|\Psi_S\rangle = \left[\left(a_0^\dagger\right)^2 + \left(a_1^\dagger\right)^2\right]|0\rangle/2$ and **e-h** an antisymmetric input state $|\Psi_{AS}\rangle = \left[\left(a_0^\dagger\right)^2 - \left(a_1^\dagger\right)^2\right]|0\rangle/2$. Best-fit corresponding simulations are shown along with experiments, optimized with input phase shifts of $\phi_S^{\text{exp.}} = 0.1\pi \pm 0.1\pi$ for the symmetric input and with $\phi_{AS}^{\text{exp.}} = 1.0\pi \pm 0.1\pi$ for the antisymmetric one. The inter-mode coupling coefficient $C = 0.45\text{cm}^{-1}$ was also determined through best-fit optimization within a range provided by a coupling-distance dependence experiment. Non-deterministic number-resolved photon detection was achieved using fiber beam-splitters.

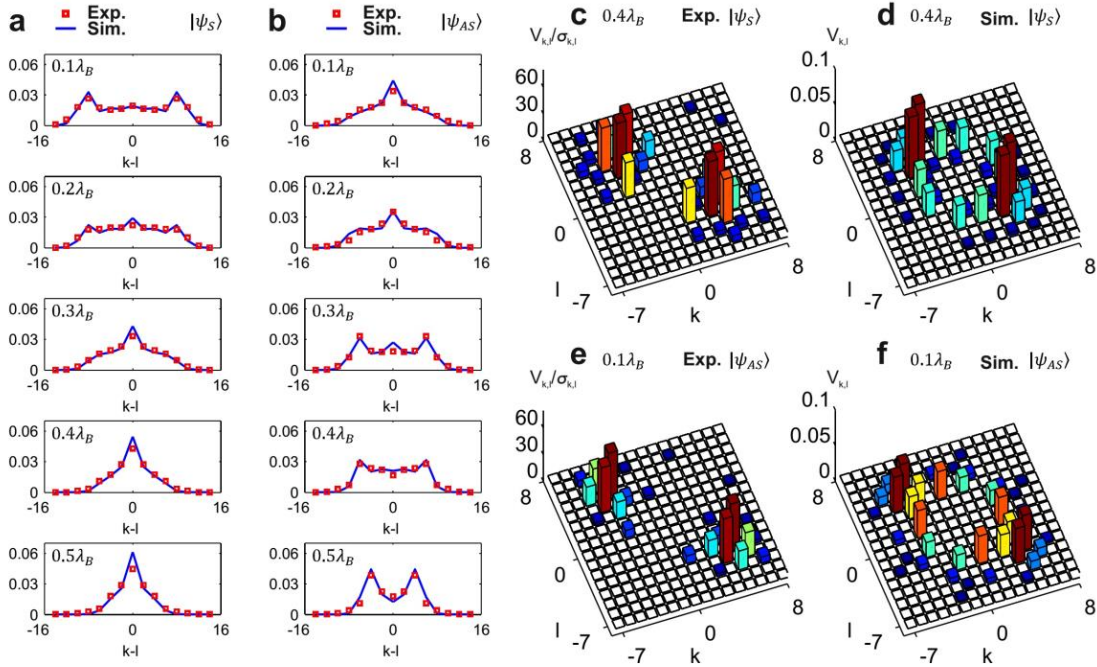

**Fig. 4.** Data analysis. **a-b,** Interparticle distance $g(k-l)$ at different discrete propagation lengths for a symmetric input state $|\Psi_S\rangle = \left[(a_0^\dagger)^2 + (a_1^\dagger)^2\right]|0\rangle/2$ and an antisymmetric input state $|\Psi_{AS}\rangle = \left[(a_0^\dagger)^2 - (a_1^\dagger)^2\right]|0\rangle/2$. **c-d,** Non-classical violations obtained with a symmetric input state at propagation length of $0.4\lambda_B$ and corresponding simulation. **e-f,** Non-classical violations obtained with an antisymmetric input state at propagation length of $0.1\lambda_B$ and corresponding simulation. A null or negative element in the inequality matrix is displayed as a blank cell.

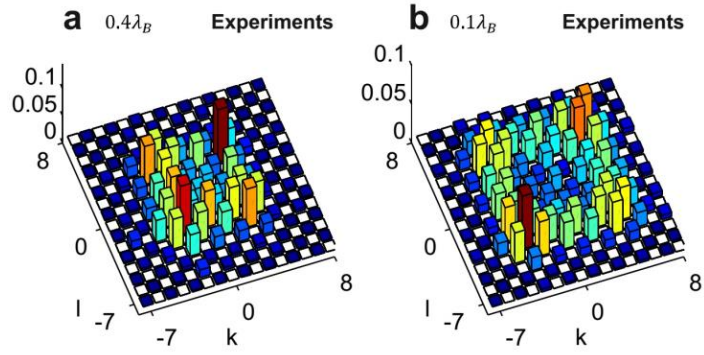

**Extended Data Fig. 1.** Experimental correlation matrices $\Gamma_{k,l}(z)$ obtained with pairs of distinguishable photons for **a** the device with symmetric input state preparation at propagation length of $0.4\lambda_B$ and **b** the device with antisymmetric input state at propagation length of $0.1\lambda_B$.

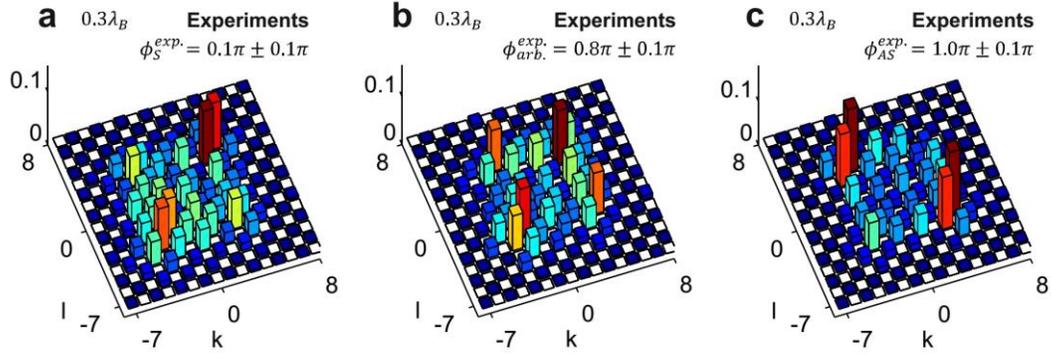

**Extended Data Fig. 2.** Experimental correlation matrices $\Gamma_{k,l}(z)$ at identical propagation length of $z = 0.3\lambda_B$ obtained through continuous tuning of the input phase shift from **a** to **b** and **c**. The phase shift of the intermediate case was determined through best-fit optimization procedure between experiments and simulations (not shown).

| Propagation length | $0.1\lambda_B$ | $0.2\lambda_B$ | $0.3\lambda_B$ | $0.4\lambda_B$ |
| --- | --- | --- | --- | --- |
| $|\Psi_S\rangle$ state | $0.950\pm0.004$ | $0.928\pm0.005$ | $0.924\pm0.002$ | $0.914\pm0.001$ |
| $|\Psi_{AS}\rangle$ state | $0.931\pm0.003$ | $0.952\pm0.002$ | $0.939\pm0.006$ | $0.936\pm0.008$ |

**Extended Data Table 1.** Similarities between experimental and simulated matrices in the cases of symmetric $|\Psi_S\rangle$ and antisymmetric $|\Psi_{AS}\rangle$ input state preparation. The errors are derived from assuming Poissonian statistics for each count and are further propagated for the calculation.

Supplementary Information for

# Bloch Oscillations of Einstein-Podolsky-Rosen States

Maxime Lebugle, Markus Gräfe, René Heilmann, Armando Perez-Leija, Stefan Nolte, and Alexander Szameit[*]

Institute of Applied Physics, Abbe Center of Photonics, Friedrich-Schiller-Universität Jena, Max-Wien-Platz 1, 07743 Jena, Germany.

* Corresponding author: alexander.szameit@uni-jena.de

## Supplementary Discussion

**Detuned directional coupler.** To be able to induce a controlled phase shift $\phi$ of the two-photon EPR state in the range $\phi \in (0, \pi)$, we take advantage of an additional degree of freedom during the fabrication process of a standard directional coupler, namely the writing speed. This parameter controls the amount of deposited laser energy through the number of pulses focused at the same spot in the glass and, therefore, directly varies the mode propagation constant. In doing so, we implement an integrated detuned Directional Coupler (detuned DC). At the single photon level, the latter can be formally described by the coupled set of equations:

$$i\frac{da_0^\dagger}{dz} = \Delta\beta a_0^\dagger - C a_1^\dagger \tag{S1}$$

$$i\frac{da_1^\dagger}{dz} = -C a_0^\dagger, \tag{S2}$$

where $a_k^\dagger$, $(k = 0,1)$ is the bosonic creation operator in waveguide $k$, $C$ is the coupling coefficient and $\Delta\beta$ the difference in the mode propagation constants. After analytical integration of coupled equations (S1) and (S2), we obtain:

$$\begin{bmatrix} a_0^\dagger(z) \\ a_1^\dagger(z) \end{bmatrix} = e^{i\Delta\beta/2} \cdot \begin{bmatrix} \cos(\kappa z) + i\dfrac{\Delta\beta}{2\kappa}\sin(\kappa z) & i\dfrac{C}{\kappa}\sin(\kappa z) \\ i\dfrac{C}{\kappa}\sin(\kappa z) & \cos(\kappa z) - i\dfrac{\Delta\beta}{2\kappa}\sin(\kappa z) \end{bmatrix} \cdot \begin{bmatrix} a_0^\dagger(0) \\ a_1^\dagger(0) \end{bmatrix}, \quad (S3)$$

where $\kappa = \sqrt{\left(\dfrac{\Delta\beta}{2}\right)^2 + C^2}$ is an effective coupling coefficient. A detuned DC behave as a balanced beam splitter if the condition

$$z = z_{BS} = \dfrac{1}{\kappa}\sin^{-1}\left(\dfrac{\kappa}{\sqrt{2}C}\right) \quad (S4)$$

is fulfilled. Remarkably, at the same propagation distance the device is able to transform a two-photon separable product state $a_0^\dagger a_1^\dagger |0\rangle$, where $|0\rangle$ is the vacuum state, into the following state:

$$\begin{aligned}
|\Psi(z_{BS})\rangle &= a_0^\dagger(z_{BS}) a_1^\dagger(z_{BS})|0\rangle \\
&= \exp\left(\dfrac{2i\Delta\beta \sin^{-1}\left(\dfrac{\kappa}{\sqrt{2}C}\right)}{\sqrt{4C^2 + \Delta\beta^2}}\right) \times \\
&\quad \left[\left(\dfrac{-\Delta\beta + iC\sqrt{4 - \dfrac{\Delta\beta^2}{C^2}}}{4C}\right)\cdot(a_0^\dagger)^2 + \left(\dfrac{+\Delta\beta + iC\sqrt{4 - \dfrac{\Delta\beta^2}{C^2}}}{4C}\right)\cdot(a_1^\dagger)^2\right]|0\rangle
\end{aligned}$$
(S5)

which is of the form $|\Psi_{EPR}\rangle = \dfrac{1}{\sqrt{2}}\left[|2_m, 0_n\rangle + e^{i\phi}|0_m, 2_n\rangle\right]$ up to a global phase factor, and thus identical to equation (2) of the main text. When prepared in such a state, the two photons are always found in the same site with a phase shift between the associated modes given by

$$\phi = \pi - 2\tan^{-1}\left(\dfrac{\sqrt{4 - \dfrac{\Delta\beta^2}{C^2}}}{\Delta\beta}\right). \quad (S6)$$

The device therefore allows for on-chip integration of arbitrary path-entangled state presenting mixed quantum statistics, or anyonic-like state. In the particular case of $\Delta\beta = 0$, one can retrieve $\phi = 0$ which corresponds to the well-known situation of a standard DC, where the two photons output the device in a symmetric wave function $|\Psi_S\rangle$. The other specific case is obtained when considering a detuning of $\Delta\beta = 2C$, generating the antisymmetric wave function $|\Psi_{AS}\rangle$, with $\phi = \pi$. Experimentally, this detuning was achieved by increasing the writing velocity to $80\,\text{mm}\cdot\text{min}^{-1}$ in one arm of the coupling section of the detuned DC, compared to $60\,\text{mm}\cdot\text{min}^{-1}$ in the other arm. The coupling length was further adjusted according to equation (S4) to obtain again a balanced device, with a splitting ratio of $50:50$. The intermediate situation for generating the arbitrary wave function $|\Psi_{arb.}\rangle$ was obtained with a writing velocity for the detuned mode of $75\,\text{mm}\cdot\text{min}^{-1}$ and led experimentally to a phase in the input state of $\phi_{arb.}^{\text{exp.}} = 0.8\pi \pm 0.1\pi$.

Every designed detuned DC was first characterized in on-chip interferometers made of one detuned DC followed by a standard DC. The bulk-optic equivalent of this scheme is the detuned Mach-Zehnder interferometer, having two arms of different lengths. Such an arrangement allows for determining the output phase shift of the detuned DC when a classical light beam is launched in one input mode, with appropriate measurement of the output ratio of the interferometer. The designed phase shift was achieved with an accuracy of about $\lambda/10$.

In the second fabrication stage the designed detuned DC was connected to the central modes of the curved waveguide array constitutive of the Bloch-oscillator. After performing the spatial correlation measurements, extensive best-fit optimization of simulated matrices was done through Root Mean Square optimization procedure, with $C$ and $\phi$ as input parameters. The results are given in the main text, further confirming the introduction of desired phase shifts in the two-photon EPR states.

**Matrices with distinguishable particles.** We consider the case of the correlations exhibited by the (antisymmetric) symmetric input state at propagation lengths (before) after the turning point of $0.25\lambda_B$ (Fig. 3d and 3e of the main text). These states are strongly correlated, implying that both photons are always found in the same branch, as a result of the cancellation of anticorrelated events. We show here that such amplitude interference is not occurring when deliberately injecting distinguishable photon pairs into the circuit, produced by adding a delay between the two particles that is longer than the wave packet coherence time. Extended Data Figure 1a-b shows the correlation matrices obtained from injecting such states in the device with symmetric input state preparation with propagation up to $0.4\lambda_B$, and with antisymmetric input state preparation with propagation up to $0.1\lambda_B$. Under these premises, the quantum state remains separable and the correlations map exhibits four peaks. Similar probabilities are therefore observed for the photon pair to gather or separate upon propagation, which is by nature a feature of separable states. Formally, it is a manifestation that time-evolved states of this kind can always be factorized, therefore preventing the building-up of non-classical correlations. No violations of the inequality $V_{k,l} = \frac{2}{3}\sqrt{\Gamma_{k,k}\Gamma_{l,l}} - \Gamma_{k,l} < 0$ are observed in this case, as was expected.